\documentclass{iaus}
\usepackage{graphicx}

\title[Brownian motion of black holes] 
{Brownian motion of black holes in stellar systems with
non-Maxwellian distribution for the stars field}

\author[Pedron and Coimbra-Ara\'ujo]
{Isabel Tamara Pedron$^1$ and Carlos H.Coimbra-Ara\'ujo$^1$}
\affiliation{$^1$Universidade Estadual do Oeste do Paran\'a \break
$85960-000$, Marechal C\^andido Rondon - PR - Brasil
\\[\affilskip]
$^2$Instituto de F\'isica Gleb Wataghin, Universidade Estadual de
Campinas\break 13083-970, Campinas - SP- Brasil }

\pubyear{2006}
\volume{238}  
\pagerange{001--999}
\date{??? and in revised form ???}
\jname{Black Holes: from Stars to Galaxies -- across the Range of Masses}
\editors{V. Karas \& G. Matt, eds.}
\begin{document}
\maketitle
\begin{abstract}
A massive black hole at the center of a dense stellar system, such
as a globular cluster or a galactic nucleus, is subject to a
random walk due gravitational encounters with nearby stars. It
behaves as a Brownian particle, since it is much more massive than
the surrounding stars and moves much more slowly than they do. If
the distribution function for the stellar velocities  is
Maxwellian, there is a exact equipartition of kinetic energy
between the black hole and the stars in the stationary state.
However,  if the distribution function deviates  from a Maxwellian
form, the strict equipartition cannot be achieved. The deviation
from equipartition is quantified in this work by applying the
Tsallis q-distribution for the stellar velocities  in a
q-isothermal stellar system  and in a generalized King model.
\end{abstract}

\firstsection 
\section{Introduction} 
 A massive black hole as
a Brownian particle conducts to a equipartition of kinetic energy
at the equilibrium, when the distribution function of stellar
velocities is Maxwellian. When it is non-Maxwellian is not
surprising to find a deviation from equipartition. This deviation
is defined as $\eta=(M\langle V^2\rangle)/(m\langle v^2\rangle)$,
where $M$ and $V$ represent the mass and velocity of the black
hole, and $m$ and $v$  masses and velocities of stars. Following
\cite[Chatterjee \etal\,(2002)]{chatt} we have
$\eta=[3\int_0^{\infty} f(r,v)vdv~\int_0^{\infty}
f(r,v)v^2dv]/[f(r,0)~\int_0^{\infty}f(r,v)v^4dv]$, with $f(r,v)$
representing the distribution function for stars field. Evidently,
for Maxwellian distribution $\eta=1$. However, astrophysical
systems are non extensive since they are subject to long range
interactions. Boltzmann-Gibbs (BG) thermostatistics is able to
predictions in extensive systems, in the sence that microscopics
interactions are short or ignored, temporal or spatial memory
effects are short range or does not exist and is valid ergodicity
in the phase space.  BG domain is enlarged by non extensive
statistical mechanics, based on Tsallis entropy (\cite[Tsallis
1988]{tsallis88}) and in the $q$-distribution function:
$p(x)\propto [1-(1-q)\beta x]^{1/(1-q)}$. Boltzmann distribution
is recovered in the $q\rightarrow 1$ limit, and all the usual
statistical mechanics as well.  A summary about mathematical
properties of these functions can be found  in \cite[Umarov \etal
(2006)]{condmat}.
\section{Extended stellar models, deviation from equipartition and results}
 The
isothermal sphere is the simplest model for spherical systems. It
 was generalized by \cite[Lima \& Souza (2005)]{lima} based on the
distribution function
$f_q(v)\propto\left[1-(1-q){v^2}/({2\sigma^2})\right]^{\frac{1}{1-q}}$
and the deviation  results in $\eta(q)=(7-5q)/[2(2-q)]$. Most
commonly used are King models, based on truncated isothermal
spheres. Extended King models were presented in \cite[Fa \& Pedron
(2001)]{pedron} and applied to fit surface brightness of the
NGC3379 and 47TUC with excellent results. The distribution
function in $q$-King models is
$f_{k_q}(\epsilon)=\rho_1(2\pi\sigma
^2)^{-3/2}\{[1+(1-q){\epsilon}/{\sigma ^2}]^{{1}/({1-q})} -1\}$
for $\epsilon > 0$ and $f_{k_q}(\epsilon) =0$ for $ \epsilon \leq
0$. Here $\epsilon$ is the
 relative energy $\epsilon =\psi- {v^2}/{2}>0$ and
$\psi=- \phi(r) +\phi_0$ the relative potential. The central
potential is  $W=\psi(0)/\sigma ^2$ and in the $\psi(0)/\sigma ^2
\rightarrow\infty$ limit the isothermal sphere is recovered. For
such model the $\eta$ value for $q<1$ and $w=-\phi(r)$ is
\begin{equation}
\eta(q,w)=\frac{3}{2}\frac{
\left[a^{-1}A^{\frac{2-q}{a}}\beta(y,1,\frac{2-q}{1-q})-w\right]\left[a^{\frac{-3}{2}}A^{\frac{5-3q}{2a}}
\beta(y,\frac{3}{2},\frac{2-q}{1-q})-\frac{2}{3}w^{3/2}\right]}
{(A^{1/a}-1)\left[a^{-5/2}A^{(7-5q)/2a}\beta(y,\frac{5}{2},\frac{2-q}{1-q})-\frac{2}{5}w^{5/2}\right]}
\end{equation}
where $\beta$ is the incomplete Beta function, $a=1-q$,
$y=\frac{w}{1/a + w}$, and  $A=1+aw$.  For $q>1$ we obtain
\begin{equation}
\eta (q,w)=\!\frac{3}{2}\frac{ \left[b^{-1} {\tilde{A}
}^{(2-q)/b}I_1-w\right]\!\!\left[b^{-\frac{3}{2}}\tilde{A}^{(5-3q)/{2b}}
I_2-\frac{2}{3}w^{3/2}\right]} {(\tilde{A}^{-1/b}-1)\left
[b^{-\frac{5}{2}}~\tilde{A}^{(7-5q)/{2b}}I_3-\frac{2}{5}w^{5/2}\right
]}
\end{equation}
where $b=q-1$, $\tilde{A}=1-bw$, and $xm=\frac{w}{1/b - w}$.
Furthermore, $I_1=\int_0^{xm}(1+x)^{-1/b} dx$,
$I_2=\int_0^{xm}(1+x)^{-1/b}x^{1/2} dx$, and
$I_3=\int_0^{xm}(1+x)^{-1/b}x^{3/2} dx$ with the conditions
$\tilde{A}=1-bw\geq 0$ and $q\leq 1+\frac{1}{w}$.

\begin{figure}[t!]
\centering
 \includegraphics[height=4.5 cm,width=6.0 cm]{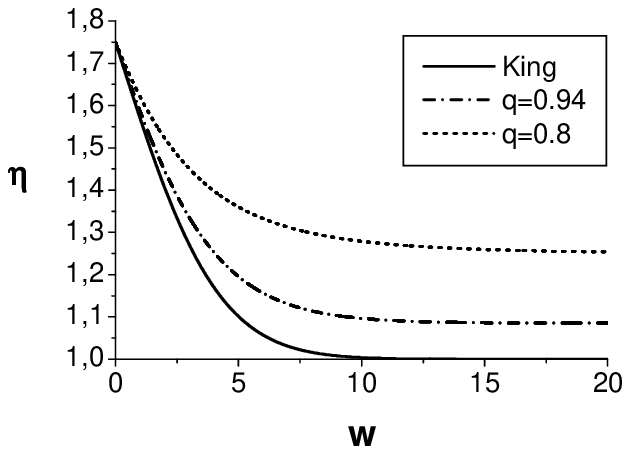}
 \includegraphics[height=4.5 cm,width=6.2 cm]{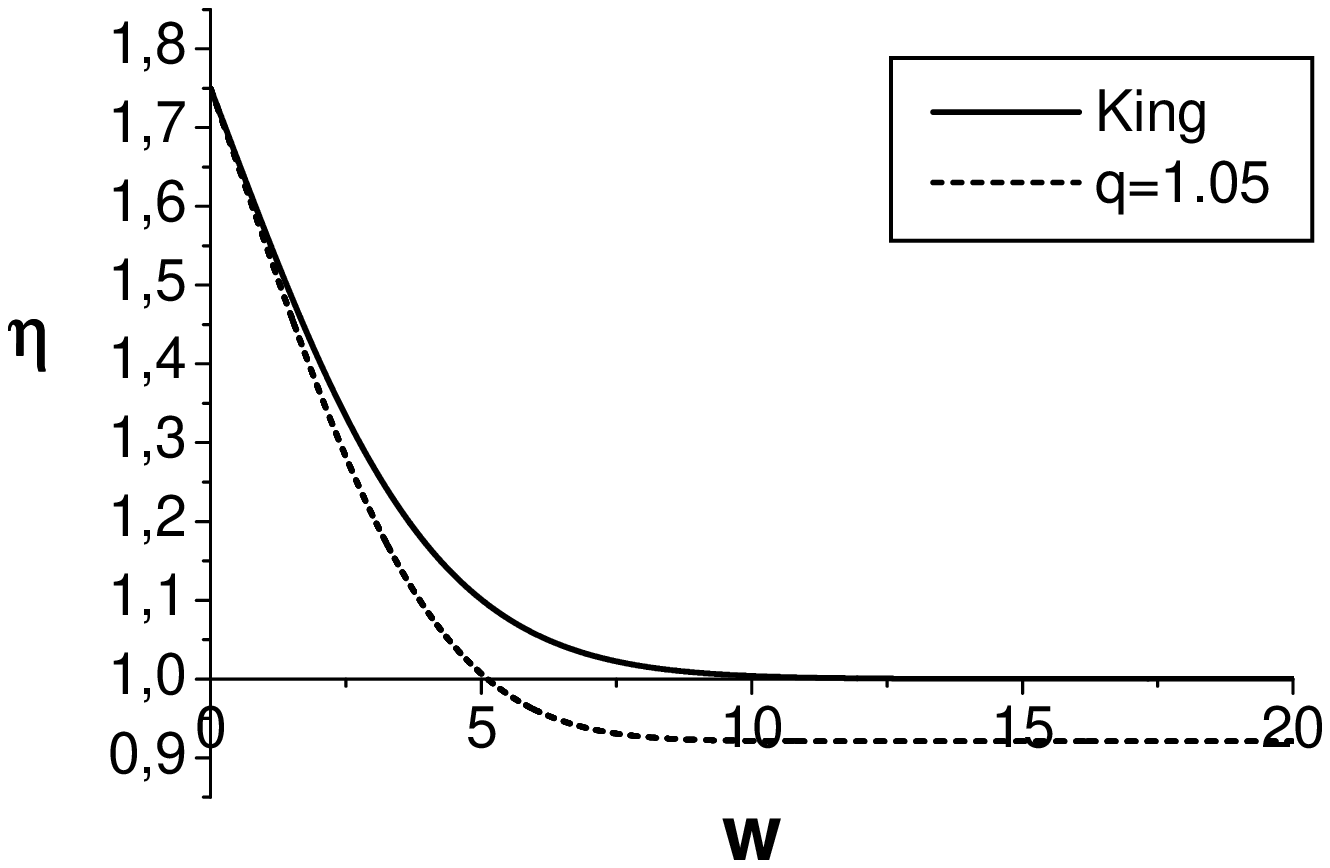}
  \caption{\small At the left, deviation for q-King model with $q=0.94$, and
$q=0.8$. The limit value is $\eta=(7-5q)/2(2-q)$. At the right,
$\eta$ for $q=1.05$.  King means $q=1$, the usual King model.
Strict equipartition implies $\eta=1$.}\label{fig1}
\end{figure}

%
%
In the Figure \ref{fig1} results  indicate that  there is a
deviation from equipartition \emph{a priori}, even at very long
time-scales. The equipartition is never achieved in both cases (
$q<1$ and $q>1$). The $q$ parameter will be, in some way, dictated
by the system.

\end{document}